\documentclass[twocolumn,showpacs,preprintnumbers,amsmath,amssymb,aps,pre]{revtex4}
\usepackage{graphicx}
\usepackage{graphics}
\usepackage{dcolumn}
\usepackage{bm}

\begin{document}

\newcommand{\tbox}[1]{\mbox{\tiny #1}}

\title{Scaling properties of discontinuous maps}

\author{J. A. M\'endez-Berm\'udez$^1$ and R. Aguilar-S\'anchez$^2$}
\affiliation{
$^1$Instituto de F\'{\i}sica, Benem\'erita Universidad Aut\'onoma de Puebla, 
Apartado Postal J-48, Puebla 72570, Mexico \\
$^2$ Facultad de Ciencias Qu\'imicas, Benem\'erita Universidad Aut\'onoma de Puebla, 
Puebla 72570, Mexico}

\date{\today}

\begin{abstract}
We study the scaling properties of discontinuous maps by analyzing 
the average value of the squared action variable $I^2$. We focus 
our study on two dynamical regimes separated by the critical value 
$K_c$ of the control parameter $K$: the slow diffusion ($K<K_c$) 
and the quasilinear diffusion ($K>K_c$) regimes. We found that the 
scaling of $I^2$ for discontinuous maps when $K\ll K_c$ and 
$K\gg K_c$ obeys the same scaling laws, in the appropriate limits, 
than Chirikov's standard map in the regimes of weak and strong 
nonlinearity, respectively. However, due to absence of KAM tori, 
we observed in both regimes that $I^2\propto nK^\beta$ for $n\gg 1$ 
(being $n$ the $n$-th iteration of the map) with $\beta\approx 5/2$ 
when $K\ll K_c$ and $\beta\approx 2$ for $K\gg K_c$.
\end{abstract}


\maketitle

\section{Introduction and model}

Chirikov's standard map (CSM), introduced in Ref.~\cite{C69}, is
an area preserving two-dimensional (2D) map for action and angle 
variables $(I,\theta)$:
\begin{eqnarray}
I_{n+1} & = & I_n + K f(\theta_n) \ , \nonumber \\
\theta_{n+1} & = & \theta_n + I_{n+1} \ , \quad \mbox{mod}-2\pi \ ,
\label{Smap}
\end{eqnarray}
where $f(\theta_n) = \sin(\theta_n)$ (due to this choice of $f(\theta)$, 
CSM is identified as a {\it continuous} map).
CSM describes the situation when nonlinear resonances are equidistant 
in phase space which corresponds to a local description of dynamical chaos
\cite{licht}. Due to this property various dynamical systems and maps can
be locally reduced to map (\ref{Smap}). Thus, CSM describes the universal 
and generic behavior of nearly-integrable Hamiltonian systems with two 
degrees of freedom having a divided phase space composed of stochastic 
motion bounded by invariant tori (also known as KAM scenario) \cite{licht}.

\begin{figure}[t]
\includegraphics[width=7cm]{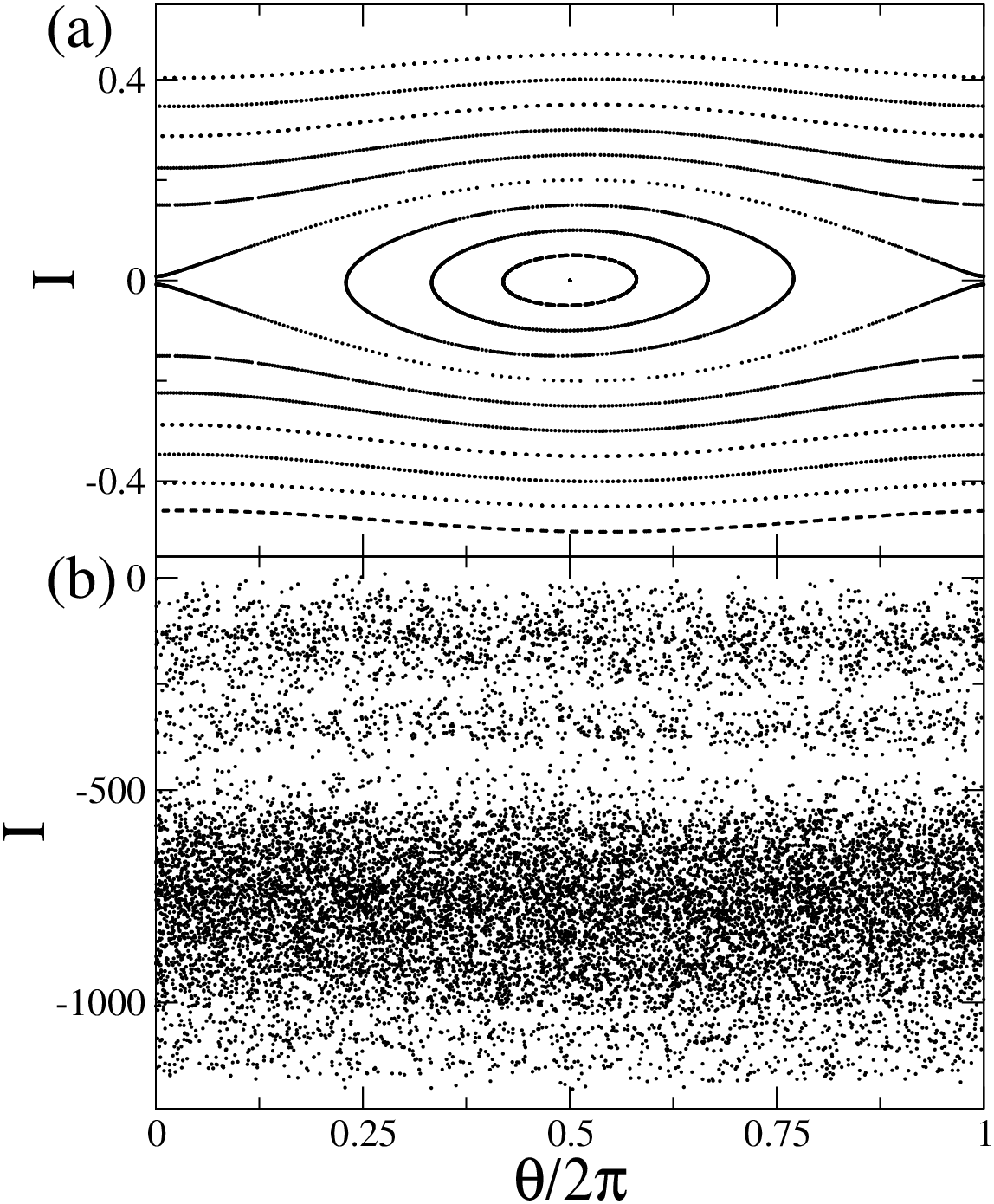}
\caption{Poincar\'e surface of section for 
CSM, Eq.~(\ref{Smap}), with (a) $K=0.01$ and 
(b) $K=10$. In (a) 20 initial conditions with $\theta_0=\pi$ and 
$I_0=[-0.5,0.5)$ were iterated $10^3$ times.
In (b) a single initial condition with $\theta_0=3$ and 
$I_0=0.01$ was iterated $3\times10^4$ times.}
\label{Fig0}
\end{figure}

CSM develops two dynamical regimes separated by the critical parameter 
$K_c$ \cite{C69,licht,C79,G79,M83,MMP84,MP85}. When $K<K_c$, regime of 
weak nonlinearity, the motion is mainly 
regular with regions of stocasticity and $I$ is bounded by KAM surfaces. 
See for example Fig.~\ref{Fig0}(a) where we present the Poincar\'e surface 
of section for CSM with $K=0.01$. Here, the value of $K$ is so small that
the Poincar\'e surface of section is equivalent to the phase portrait of 
a one-dimensional pendulum.
At $K=K_c$, the last KAM curve is destroyed 
and the transition to global stocasticity takes place. 
Then, for $K>K_c$, regime of strong nonlinearity, $I$ becomes unbounded 
and increases diffusively. See for example the Poincar\'e map of 
Fig.~\ref{Fig0}(b) where a 
single trajectory has been iterated $3\times10^4$ times.

Even though CSM describes the universal behavior of area-preserving 
continuous maps, another class of Hamiltonian dynamical 
systems is represented by the {\it discontinuous} map \cite{B98}:
\begin{eqnarray}
I_{n+1} & = & I_n + K f(\theta_n) \ , \nonumber \\
\theta_{n+1} & = & \theta_n + T I_{n+1} \ , \quad \mbox{mod}-2\pi \ ,
\label{map}
\end{eqnarray}
where $f(\theta_n) = \sin(\theta_n) \mbox{sgn}[\cos(\theta_n)]$. 
Examples of physical systems described by discontinuous maps are
2D billiard models like the stadium billiard \cite{stadium1,stadium2}
and polygonal billiards \cite{poly1,poly2}. The origin of the
discontinuity in map (\ref{map}) are the sudden translations of
the action under the system dynamics.

\begin{figure}[t]
\includegraphics[width=7cm]{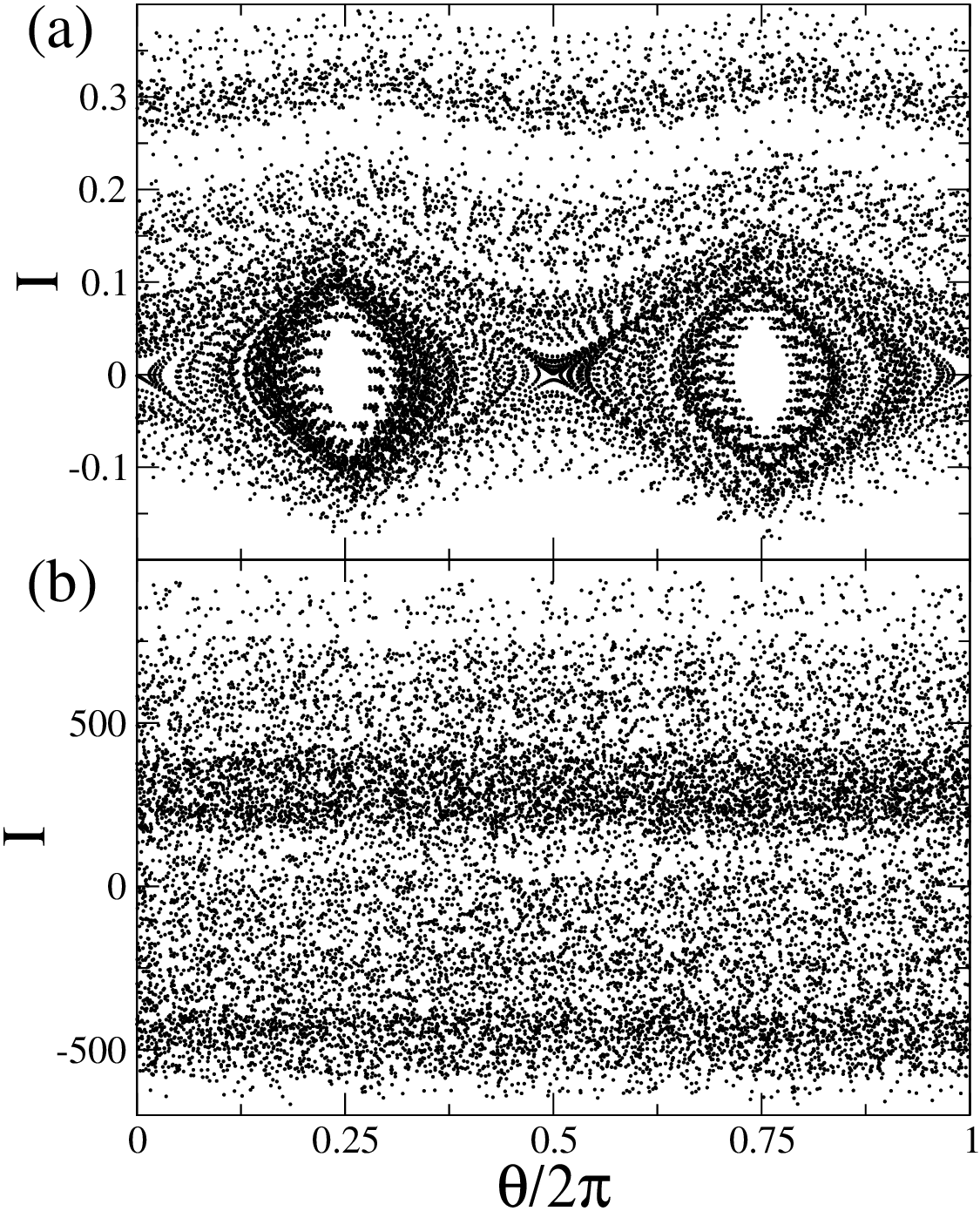}
\caption{Poincar\'e surface of section for 
the discontinuous map of Eq.~(\ref{map}) with (a) $K=0.01$ and 
(b) $K=10$. $T=1$. A single initial condition with $\theta_0=3$ 
and $I_0=0.01$ was iterated $3\times10^4$ times.}
\label{Fig1}
\end{figure}

As well as CSM, map (\ref{map}) is known to have 
two different dynamical regimes, however both diffusive, delimited by 
the critical value $K_c=1/T$ \cite{B98}.
The regimes $K<K_c$ and $K>K_c$ are known as slow 
diffusion and quasilinear diffusion regimes, respectively. 
As an example of the dynamics of map (\ref{map}), in 
Fig.~\ref{Fig1} we show typical Poincar\'e surface of sections in 
both regimes (for comparison purposes we have used the same values
of $K$ as in Fig.~\ref{Fig0} for CSM). 
On the one hand, as can be observed by contrasting Figs.~\ref{Fig0}(a) 
and \ref{Fig1}(a), the main difference between CSM and map 
(\ref{map}) is that for $K<K_c$ the later does not show 
regular behavior. In fact, due to the discontinuities of 
$f(\theta)$ in map (\ref{map}), KAM theorem is not satisfied
and map (\ref{map}) does not develop the KAM scenario. 
Since for any $K\ne 0$
the dynamics of map (\ref{map}) is diffusive, a single trajectory
can explore the entire phase space. However, in the slow 
diffusion regime the dynamics is far from being stochastic due to 
the sticking of trajectories along cantori (fragments of KAM invariant 
tori), see Fig.~\ref{Fig1}(a). On the other hand, for $K>K_c$ map 
(\ref{map}) shows diffusion similar to that of CSM, compare 
Figs.~\ref{Fig0}(b) and \ref{Fig1}(b).
We want to add that independently of the value of $K\ne 0$, map 
(\ref{map}) has five period-one fixed points at $I=0$ and 
$\theta=[0,\pi/2,\pi,3\pi/2,2\pi]$.

In particular, in Ref.~\cite{LS07} a scaling analysis of CSM was performed 
by studying the average value of $I^2$ as a 
function of $K$ and the $n$-th iteration of the map. There, the following
scaling law was reported:
\begin{eqnarray} 
I^2\propto n^\alpha K^\beta \ ;
\label{scaling} 
\end{eqnarray}
where $\alpha\approx 2$ for $K\ll K_c$ and small $n$
while $\alpha\approx 1$ for $K\gg K_c$ and large $n$, 
with $\beta\approx 2$ 
in both cases. The scaling (\ref{scaling}) has also been validated 
for several dynamical systems represented by the standard map
such as the Fermi-Ulam model \cite{ulam1,ulam2,ulam3,ulam4,ulam5}, 
time-dependent potential wells \cite{well}, and
waveguide billiards \cite{ulam5,waveguide}; among others 
\cite{other1,other2}.

Although map (\ref{map}) has the same structure than CSM, 
a systematic investigation of the scaling properties of $I^2$ for 
discontinuous maps is not available in the literature. Thus, in this 
paper we undertake this task. For this purpose, here we study 
the properties of the map of Eq.~(\ref{map}) by analyzing the 
scaling of the average value of the squared action variable 
$I^2$ as a function of $n$, $K$, and $I_0$. 
We choose the scaling approach to $I^2$ reported in Ref.~\cite{LS07} 
because of the similarity of maps (\ref{Smap}) and (\ref{map}).
Moreover, since map (\ref{map}) shows diffusion in both dynamical 
regimes ($K<K_c$ and $K>K_c$), we expect the scaling (\ref{scaling}) 
to be valid for discontinuous maps when diffusion is present with 
scaling exponents $\alpha$ and $\beta$ to be determined.

\begin{figure}[t]
\includegraphics[width=8.5cm]{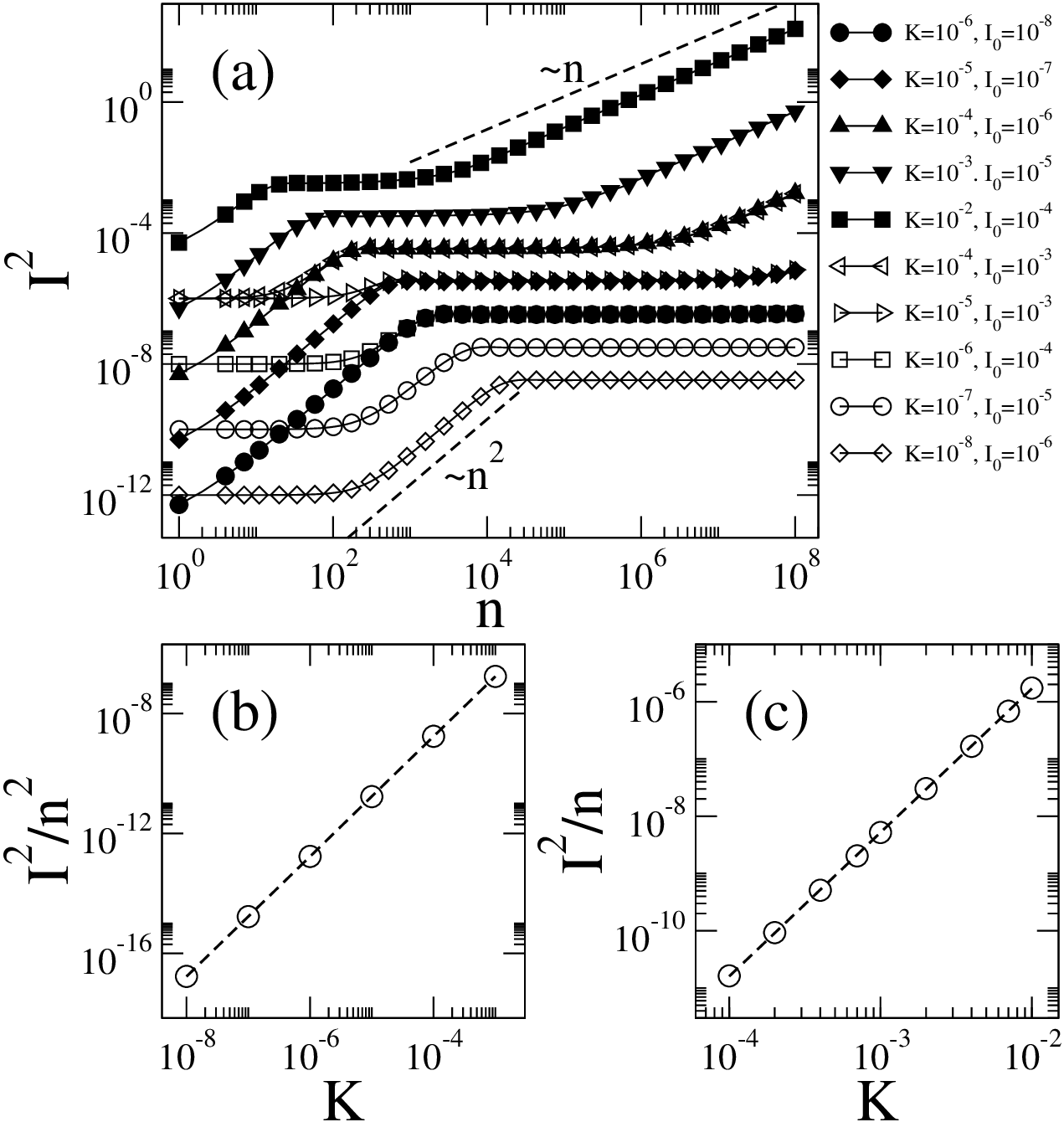}
\caption{(a) $I^2$ as a function of $n$ in the 
slow diffusion regime ($K \ll 1$). Full symbols (open symbols) correspond to $I_0 \ll K$ 
($I_0 \gg K$). Each curve is the average over 1000
trajectories having initial random phases in the interval 
$0<\theta_0<2\pi$. The dashed lines proportional to $n$ and
$n^2$ are plotted to guide the eye.
(b) [(c)] $I^2/n^2$ [$I^2/n$] as a function of $K$ for 
$n<n_{\tbox{cr}}^{(1)}$ [$n>n_{\tbox{cr}}^{(2)}$].
The dashed line equal to $0.17K^2$ [$0.17K^{5/2}$] is the best 
power law-fit to the data.}
\label{Fig2a}
\end{figure}
\begin{figure}[t]
\includegraphics[width=8cm]{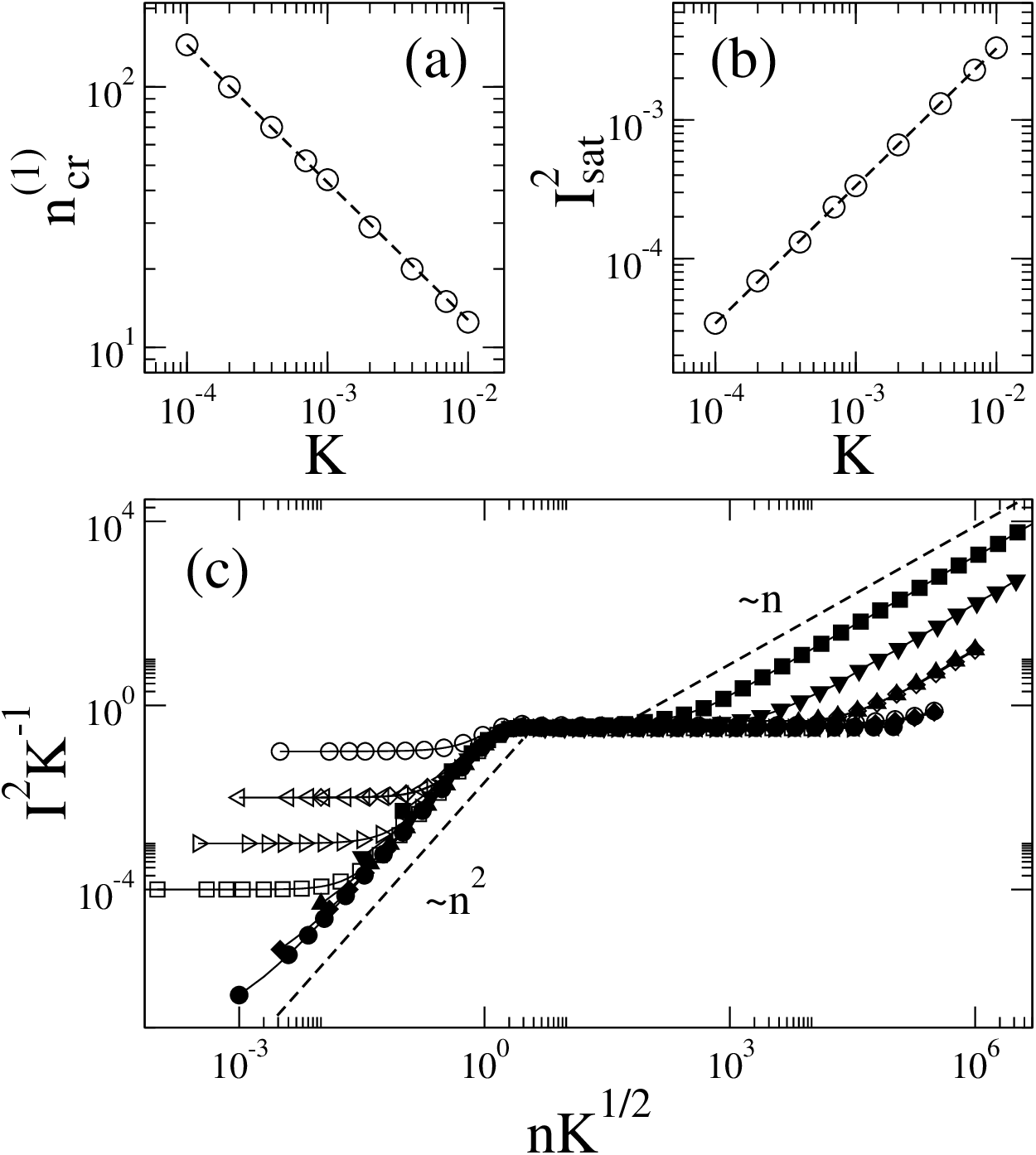}
\caption{(a) [(b)] $n_{\tbox{cr}}^{(1)}$ [$I^2_{\tbox{sat}}$] as a 
function of $K$. The dashed line equal to $1.1K^{-1/2}$ [$0.31K$] 
is the best power-law fit to the data.
(c) Scaled curves $I^2K^{-1}$ as a function of $nK^{1/2}$ in the 
slow diffusion regime ($K \ll 1$). Full symbols (open symbols) correspond to $I_0 \ll K$ 
($I_0 \gg K$). Same data as in Fig.~\ref{Fig2a}. The dashed 
lines show that $I^2\propto n^2$ for $n<n_{\tbox{cr}}^{(1)}$ 
while $I^2\propto n$ for $n>n_{\tbox{cr}}^{(2)}$.}
\label{Fig2b}
\end{figure}

\section{Results}

We compute $I^2$ for map (\ref{map}) following two steps 
\cite{LS07}: First we calculate the average squared action 
over the orbit associated with the initial condition $j$ as
\[
\left< I^2_{n,j} \right> = \frac{1}{n+1} \sum^n_{i=0} I^2_{i,j} \ ,
\]  
where $i$ refers to the $i$-th iteration of the map. Then, 
the average value of $I^2$ is defined as the average over 
$M$ independent realizations of the map (by randomly choosing 
values of $\theta_0$):
\begin{equation}
I^2(n,K,I_0) = \frac{1}{M} \sum^M_{j=1} \left< I^2_{n,j} \right> \ .
\label{I2}
\end{equation}
In the following, without lost of generality, we set $T=1$.

\subsection{Slow diffusion regime}

In Fig.~\ref{Fig2a}(a) we present $I^2$ as a function of $n$ in the 
slow diffusion regime ($K \ll 1$) for several combinations of $K$ 
and $I_0$. In fact, $I^2$ is always an increasing function of $n$, 
however its growth is marginal in some iteration intervals producing 
plateaus in the curves $I^2$ vs $n$.

For $I_0 \ll K$, see full symbols in Fig.~\ref{Fig2a}(a), $I^2$ grows 
up to a crossover iteration number $n_{\tbox{cr}}^{(1)}$.
When $n_{\tbox{cr}}^{(1)}<n<n_{\tbox{cr}}^{(2)}$ the trajectories
wander around the period-one fixed points of the map making the growth 
of $I^2$ negligible; that is, $I^2$ becomes almost constant. 
We call this constant $I^2_{\tbox{sat}}$. 
Then, for $n>n_{\tbox{cr}}^{(2)}$, the trajectories scape from the 
influence of the period-one fixed points and $I^2$ starts to increase again.

In Fig.~\ref{Fig2a}(a) we also explore the case $I_0 \gg K$, see open symbols. 
During the first few iteration steps, since $K$ is small as compared to 
$I_0$, $I^2$ does not increase significantly as a function of $n$; so, $I^2$
remains approximately equal to $I_0^2$ up to a crossover iteration number 
$n_{\tbox{cr}}^{(0)}$.
For $n>n_{\tbox{cr}}^{(0)}$, $I^2$ 
follows the same panorama when increasing $n$ as it does in the case 
$I_0 \ll K$: it grows up to $n_{\tbox{cr}}^{(1)}$, 
then it becomes approximately equal to $I^2_{\tbox{sat}}$ up to 
$n_{\tbox{cr}}^{(2)}$, and finally it grows again. 

Then, based on Fig.~\ref{Fig2a}(a), we postulate the following scaling
relations:
\begin{equation}
I^2(n,K) \propto n^\alpha K^\beta 
\label{scaling1}
\end{equation}
for $n<n_{\tbox{cr}}^{(1)}$ and $n>n_{\tbox{cr}}^{(2)}$, with
\begin{equation}
n_{\tbox{cr}}^{(1)}(K) \propto K^{\gamma_1}
\label{scaling2} 
\end{equation}
and
\begin{equation}
n_{\tbox{cr}}^{(2)}(K) \propto K^{\gamma_2} \ ;
\label{scaling3} 
\end{equation}
in addition
\begin{equation}
I^2_{\tbox{sat}}(K) \propto K^\delta \ .
\label{scaling4}
\end{equation}
Also, from Fig.~\ref{Fig2a}(a), we concluded that 
$n_{\tbox{cr}}^{(0)} = \mbox{const.} \approx 215$.
Below, we present a detailed analysis that allows us to obtain the scaling 
exponents $\alpha$, $\beta$, $\gamma_{1,2}$, and $\delta$. 

By performing power-law fittings to the growth regimes of $I^2$, we determined 
that $\alpha\approx 2$ for $n<n_{\tbox{cr}}^{(1)}$ and $\alpha\approx 1$ 
when $n>n_{\tbox{cr}}^{(2)}$. See dashed lines in Fig.~\ref{Fig2a}(a).
Once we know the exponents $\alpha$ we can extract the exponents $\beta$. 
To this end, in Figs.~\ref{Fig2a}(b) and \ref{Fig2a}(c) we plot $I^2/n^2$ for 
$n<n_{\tbox{cr}}^{(1)}$ and $I^2/n$ for $n>n_{\tbox{cr}}^{(2)}$, respectively, 
as a function of $K$.
The dashed lines, equal to $0.17K^2$ and $0.17K^{5/2}$, which are the 
best power-law fits to the data, prove that $\beta\approx 2$ for 
$n<n_{\tbox{cr}}^{(1)}$ and $\beta\approx 5/2$ when $n>n_{\tbox{cr}}^{(2)}$.
In fact, the dependence $I^2\propto K^{5/2}$ for $n>n_{\tbox{cr}}^{(2)}$ is 
not surprising since theoretical results for the saw-tooth
map \cite{saw} as well as numerical computations on the stadium map 
\cite{stadium1} (both discontinuous maps) show that 
$I^2\propto K^{5/2}$ when $K<K_c$ for large $n$.
More precisely, for $K<K_c$ the dynamics of map (\ref{map}) is 
diffusive (after the transient time $n_{\tbox{cr}}^{(2)}$) with 
diffusion rate $D=D_0 K^{5/2}\sqrt{T}$ \cite{B98}, 
where $D=\lim_{n\to \infty} \left< I^2(n) \right> /n$ and the 
average $\left< \cdots \right>$ is performed over an ensemble 
of trajectories with the same initial action $I_0$ and random 
initial phases $\theta_0$. Here, $D_0\approx 0.4$ is the constant proper 
of the choice of $f(\theta)$ we made \cite{B98}.

Then, in Figs.~\ref{Fig2b}(a) and \ref{Fig2b}(b) we show 
$n_{\tbox{cr}}^{(1)}$ and $I^2_{\tbox{sat}}$ as a function of $K$,
respectively. We computed $n_{\tbox{cr}}^{(1)}$ as the intersection
of a power-law fitting curve $I^2\propto n^2$ for $n<n_{\tbox{cr}}^{(1)}$
with the constant curve $I^2=I^2_{\tbox{sat}}$.
The dashed lines in Figs.~\ref{Fig2b}(a) and \ref{Fig2b}(b) equal to 
$1.1K^{-1/2}$ and $0.31K$, respectively, lead to $\gamma_1\approx -1/2$
and $\delta\approx 1$.
As a consequence of the scalings above, in Fig.~\ref{Fig2b}(c) we present 
the scaled curves $I^2K^{-1}$ as a function of $nK^{1/2}$ showing the 
collapse of $I^2_{\tbox{sat}}$ and $n_{\tbox{cr}}^{(1)}$.

\begin{figure}[t]
\includegraphics[width=8cm]{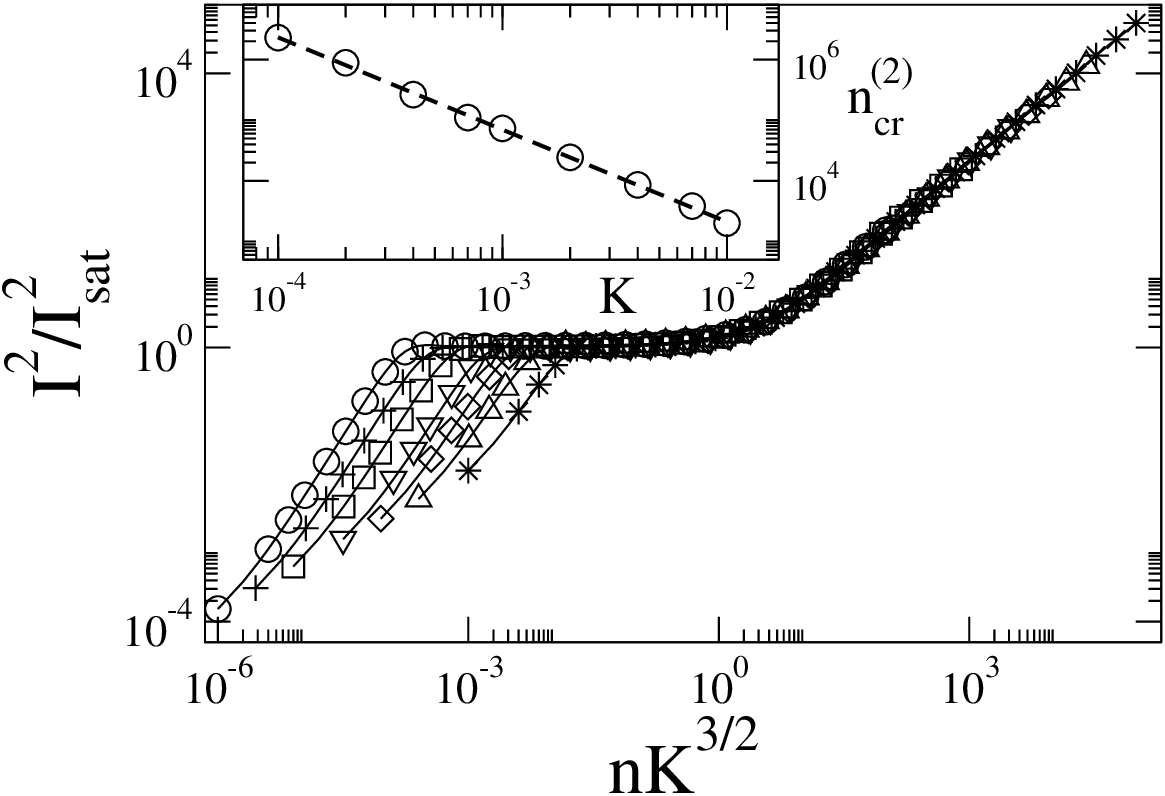}
\caption{$I^2/I^2_{\tbox{sat}}$ as a function of $nK^{3/2}$ 
in the slow diffusion regime ($K \ll 1$). From left to right:
$K=10^{-4}$, $2\times10^{-4}$, $4\times10^{-4}$, $10^{-3}$,
$2\times10^{-3}$, $4\times10^{-3}$, and $10^{-2}$. $I_0=K/100$. 
Inset: $n_{\tbox{cr}}^{(2)}$ as a function of $K$. The dashed 
line equal to $1.9K^{-3/2}$ is the best power law-fit to the data.}
\label{Fig3}
\end{figure}

We want to stress that the scaling of $I^2$ for the discontinuous map 
of Eq.~(\ref{map}) in the slow diffusion regime obeys the same scaling 
laws than CSM in the regime of weak nonlinearity, see \cite{LS07}, 
except for the appearance of the second crossover iteration number 
$n_{\tbox{cr}}^{(2)}$. To study the dependence of $n_{\tbox{cr}}^{(2)}$ 
on $K$, in Fig.~\ref{Fig3}(Inset) we plot $n_{\tbox{cr}}^{(2)}$ vs $K$. 
The power-law fitting of the data leads to $\gamma_2 \approx -3/2$ and 
a proportionality constant $\approx 1.9$. Then, in the main panel of 
Fig.~\ref{Fig3} we show that the curves $I^2/I^2_{\tbox{sat}}$ are 
properly scaled, for $n>n_{\tbox{cr}}^{(2)}$, when plotting them as a 
function of $nK^{3/2}$. The behavior $I^2\propto n$ for 
$n>n_{\tbox{cr}}^{(2)}$ should be expected in map (\ref{map}) since 
here, in contrast to CSM with $K<K_c$, the movement is not bounded by 
KAM tori and particles can diffuse along the phase space cylinder without 
limit.

\begin{figure}[t]
\includegraphics[width=8cm]{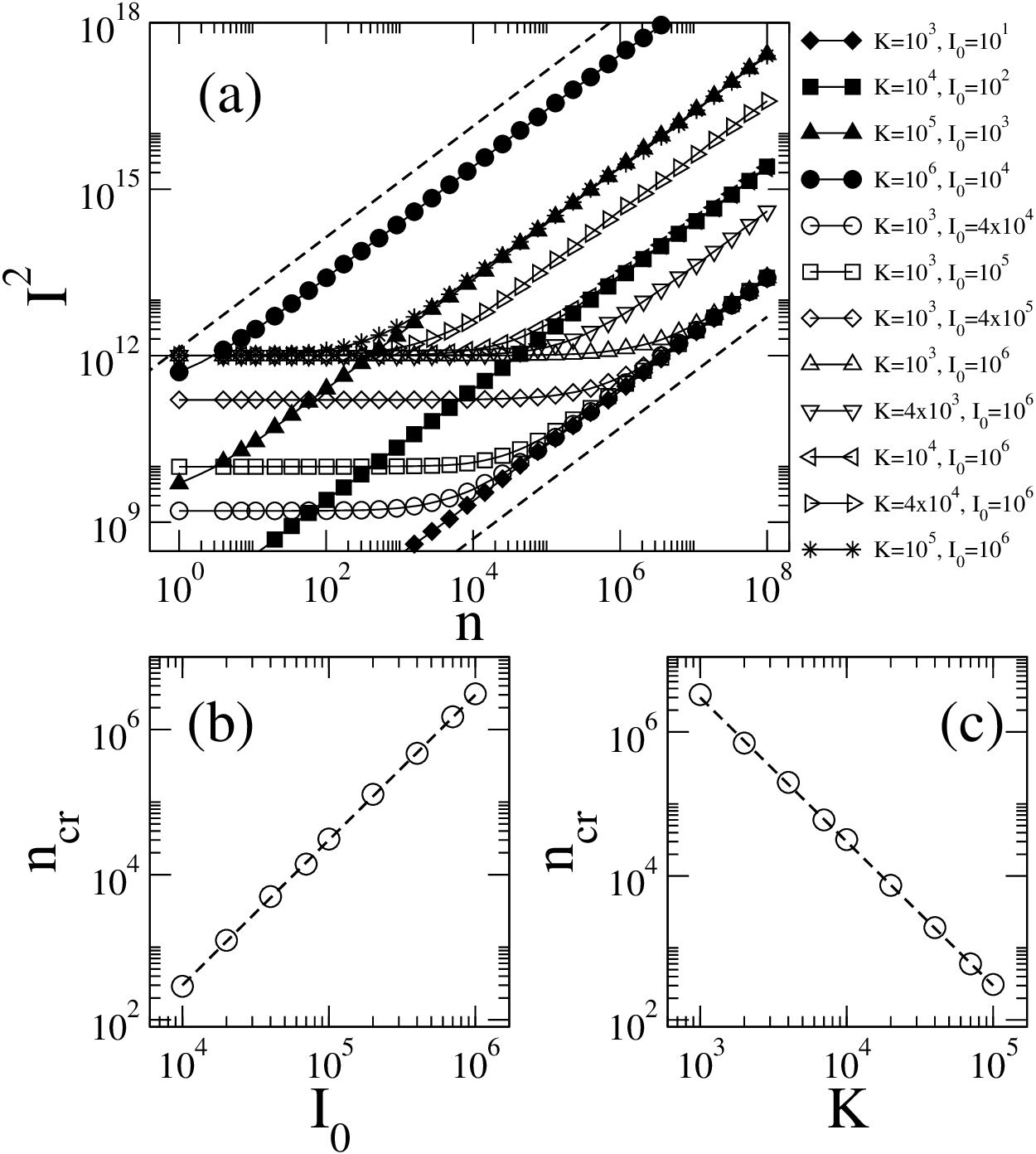}
\caption{(a) $I^2$ as a function of $n$ in 
the quasilinear diffusion regime ($K \gg 1$). Open symbols (full symbols) 
correspond to $I_0 \gg K$ ($I_0 \ll K$). Each curve is the average 
over 1000 trajectories having initial random phases in the 
interval $0<\theta_0<2\pi$. The dashed lines, plotted to guide 
the eye, are proportional to $n$. (b) [(c)] $n_{\tbox{cr}}$ as a 
function of $I_0$ [$K$] for $K=10^3$ [$I_0=10^6$]. The dashed 
line equal to $3I_0^2$ [$3.1K^{-2}$] is the best power 
law-fit to the data.}
\label{Fig4a}
\end{figure}
\begin{figure}[t]
\includegraphics[width=8cm]{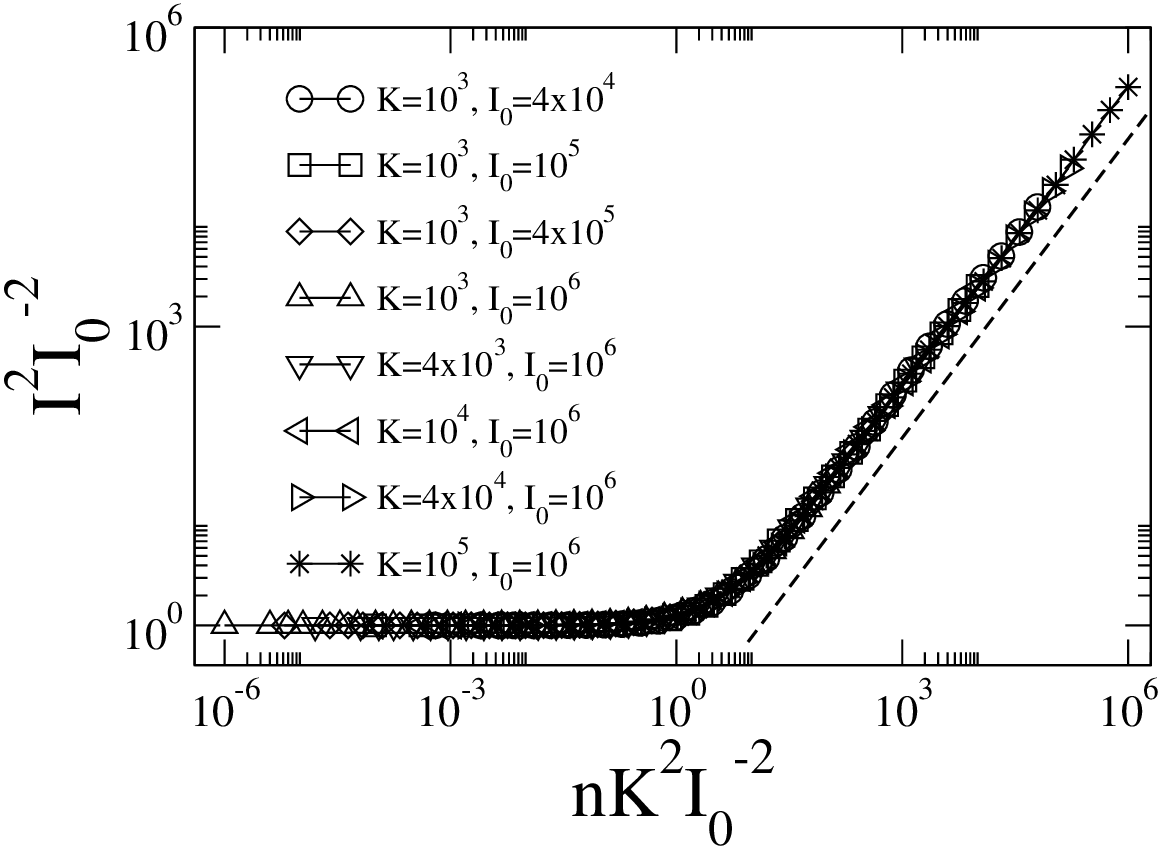}
\caption{Scaled curves $I^2I_0^{-2}$ as a function of 
$nK^2I_0^{-2}$ for $I_0 \gg K$. The dashed line shows that 
$I^2\propto n$ for $n>n_{\tbox{cr}}$.}
\label{Fig4b}
\end{figure}

\subsection{Quasilinear diffusion regime}

In Fig.~\ref{Fig4a}(a) we present $I^2$ as a function of $n$ in the 
quasilinear diffusion regime ($K \gg 1$) for several combinations of 
$K$ and $I_0$.

For $I_0 \ll K$, $I^2$ grows proportional to $n$ for all $n$. See 
full symbols in Fig.~\ref{Fig4a}(a). For $I_0 \gg K$, $I^2$ as a 
function of $n$ is almost constant and approximately equal to $I_0^2$ up to a crossover 
iteration number $n_{\tbox{cr}}$. Then, when $n>n_{\tbox{cr}}$, $I^2$ increases
proportional to $n$. See open symbols in Fig.~\ref{Fig4a}(a).
This behavior for $I^2$ is completely equivalent to that for 
CSM in the strong nonlinearity regime \cite{LS07}.
That is, the scaling given in Eq.~(\ref{scaling}) is valid for
$n>n_{\tbox{cr}}$ with $\alpha\approx 1$ and $\beta\approx 2$.
This is consistent with the random phase approximation \cite{C79} 
that predicts, for $K>K_c$, diffusive motion along the $I$ direction 
with a diffusion rate $D = K^2/2$.
Moreover, we observed that the crossover iteration number $n_{\tbox{cr}}$ 
scales as
\begin{equation}
n_{\tbox{cr}}(K,I_0) \propto K^{\gamma_3}I_0^{\gamma_4} \ .
\label{scaling5} 
\end{equation}
To get the exponents $\gamma_{3,4}$ in the scaling relation 
above in Figs.~\ref{Fig4a}(b) and \ref{Fig4a}(c) we plot: 
(i) $n_{\tbox{cr}}$ as a function of $I_0$ for fixed $K$ and
(ii) $n_{\tbox{cr}}$ as a function of $K$ for fixed $I_0$, respectively. 
Using power-law fittings, see Figs.~\ref{Fig4a}(b) and \ref{Fig4a}(c), 
we found that $n_{\tbox{cr}}\propto K^{-2}I_0^2$  
with a proportionality constant $\approx 3$. Thus, we concluded that
$\gamma_3\approx -2$ and $\gamma_4\approx 2$. Finally, from scaling 
(\ref{scaling5}), in Fig.~\ref{Fig4b} we show that all curves 
$I^2I_0^{-2}$ as a function of $nK^2I_0^{-2}$ colapse into a single one.

\section{Conclusions}

We have studied the scaling properties of the action variable $I$ 
for the discontinuous map of Eq.~(\ref{map}). 
We focus on the slow diffusion ($K<K_c$) and the quasilinear 
diffusion ($K>K_c$) regimes, being $K_c=1/T$.
We found that the scaling of $I^2$ for map (\ref{map}) when 
$K\ll K_c$ and $K\gg K_c$ obey the same scaling laws than CSM 
in the regimes of weak and strong nonlinearity 
\cite{LS07}, respectively. Except for that in the slow diffusion
regime, due to the absence of KAM tori to bound the motion,
$I^2\propto n$ for large enough $n$. 
Also, we conclude that the scaling $I^2\propto n^\alpha K^\beta$ applies to
discontinuous maps with 
\begin{itemize}
\item[(i)] $\alpha\approx 2$ and $\beta\approx 2$ 
for $K\ll K_c$ and small $n$;
\item[(ii)] $\alpha\approx 1$ and $\beta\approx 5/2$ 
for $K\ll K_c$ and large $n$; and 
\item[(iii)] $\alpha\approx 1$ and $\beta\approx 2$ 
for $K\gg K_c$ and large $n$.
\end{itemize} 
Our results are summarized in Table~\ref{Table1}.

\begin{table}[t]
  \centering
  \begin{tabular}{|l|c|c|c|c|} \\ \hline \hline
    & $K\ll 1$    & $K\ll 1$   & $K\gg 1$ & $K\gg 1$ \\ 
    & $I_0\ll K$  & $I_0\gg K$ & $I_0\ll K$ & $I_0\gg K$ \\ \hline \hline
$I^2\approx I^2_0$ & --- & $n<n_{\tbox{cr}}^{(0)}$ & --- & $n<n_{\tbox{cr}}$ \\
$I^2\propto n^2K^2$ & $n<n_{\tbox{cr}}^{(1)}$ &  
$n_{\tbox{cr}}^{(0)}<n<n_{\tbox{cr}}^{(1)}$ & --- & --- \\
$I^2\approx I^2_{\tbox{sat}}$ & $n_{\tbox{cr}}^{(1)}<n<n_{\tbox{cr}}^{(2)}$  & $n_{\tbox{cr}}^{(1)}<n<n_{\tbox{cr}}^{(2)}$ & --- & --- \\
$I^2\propto nK^{5/2}$ & $n>n_{\tbox{cr}}^{(2)}$ & $n>n_{\tbox{cr}}^{(2)}$ & 
--- & --- \\ 
$I^2\propto nK^2$ & --- & --- & $n>n_{\tbox{cr}}$ & $n>n_{\tbox{cr}}$ 
\\ \hline \hline
  \end{tabular}
\caption{Behavior of $I^2$ in the slow diffusion ($K\ll 1$) and the 
quasilinear diffusion ($K\gg 1$) regimes. We have found that
$I^2_{\tbox{sat}}\approx 0.31K$, $n_{\tbox{cr}}^{(0)}\approx 215$, 
$n_{\tbox{cr}}^{(1)}\approx 1.1K^{-1/2}$,
$n_{\tbox{cr}}^{(2)}\approx 1.9K^{-3/2}$, and 
$n_{\tbox{cr}}\approx 3I_0^2K^{-2}$.}
\label{Table1}
\end{table}

\acknowledgments

This work was partially supported by VIEP-BUAP 
(grants MEBJ-EXC10-I and SARA-NAT10-I) and PROMEP
(grants 103.5/09/4194 and 103.5/10/8442), Mexico.

\end{document}